\documentclass{article}
\usepackage{amsmath, amssymb, amsfonts}
\title{ Geodesic deviation in a nonlinear gravitational wave spacetime } 
\author{Hristu Culetu, \\Ovidius University, Dept.of Physics and Electronics, \\B-dul Mamaia 124, 900527 Constanta, Romania, \\e-mail : hculetu@yahoo.com}

\begin{document}
\numberwithin{equation}{section}
\pagenumbering{arabic}
\maketitle
\newcommand{\fv}{\boldsymbol{f}}
\newcommand{\tv}{\boldsymbol{t}}
\newcommand{\gv}{\boldsymbol{g}}
\newcommand{\OV}{\boldsymbol{O}}
\newcommand{\wv}{\boldsymbol{w}}
\newcommand{\WV}{\boldsymbol{W}}
\newcommand{\NV}{\boldsymbol{N}}
\newcommand{\hv}{\boldsymbol{h}}
\newcommand{\yv}{\boldsymbol{y}}
\newcommand{\RE}{\textrm{Re}}
\newcommand{\IM}{\textrm{Im}}
\newcommand{\rot}{\textrm{rot}}
\newcommand{\dv}{\boldsymbol{d}}
\newcommand{\grad}{\textrm{grad}}
\newcommand{\Tr}{\textrm{Tr}}
\newcommand{\ua}{\uparrow}
\newcommand{\da}{\downarrow}
\newcommand{\ct}{\textrm{const}}
\newcommand{\xv}{\boldsymbol{x}}
\newcommand{\mv}{\boldsymbol{m}}
\newcommand{\rv}{\boldsymbol{r}}
\newcommand{\kv}{\boldsymbol{k}}
\newcommand{\VE}{\boldsymbol{V}}
\newcommand{\sv}{\boldsymbol{s}}
\newcommand{\RV}{\boldsymbol{R}}
\newcommand{\pv}{\boldsymbol{p}}
\newcommand{\PV}{\boldsymbol{P}}
\newcommand{\EV}{\boldsymbol{E}}
\newcommand{\DV}{\boldsymbol{D}}
\newcommand{\BV}{\boldsymbol{B}}
\newcommand{\HV}{\boldsymbol{H}}
\newcommand{\MV}{\boldsymbol{M}}
\newcommand{\be}{\begin{equation}}
\newcommand{\ee}{\end{equation}}
\newcommand{\ba}{\begin{eqnarray}}
\newcommand{\ea}{\end{eqnarray}}
\newcommand{\bq}{\begin{eqnarray*}}
\newcommand{\eq}{\end{eqnarray*}}
\newcommand{\pa}{\partial}
\newcommand{\f}{\frac}
\newcommand{\FV}{\boldsymbol{F}}
\newcommand{\ve}{\boldsymbol{v}}
\newcommand{\AV}{\boldsymbol{A}}
\newcommand{\jv}{\boldsymbol{j}}
\newcommand{\LV}{\boldsymbol{L}}
\newcommand{\SV}{\boldsymbol{S}}
\newcommand{\av}{\boldsymbol{a}}
\newcommand{\qv}{\boldsymbol{q}}
\newcommand{\QV}{\boldsymbol{Q}}
\newcommand{\ev}{\boldsymbol{e}}
\newcommand{\uv}{\boldsymbol{u}}
\newcommand{\KV}{\boldsymbol{K}}
\newcommand{\ro}{\boldsymbol{\rho}}
\newcommand{\si}{\boldsymbol{\sigma}}
\newcommand{\thv}{\boldsymbol{\theta}}
\newcommand{\bv}{\boldsymbol{b}}
\newcommand{\JV}{\boldsymbol{J}}
\newcommand{\nv}{\boldsymbol{n}}
\newcommand{\lv}{\boldsymbol{l}}
\newcommand{\om}{\boldsymbol{\omega}}
\newcommand{\Om}{\boldsymbol{\Omega}}
\newcommand{\Piv}{\boldsymbol{\Pi}}
\newcommand{\UV}{\boldsymbol{U}}
\newcommand{\iv}{\boldsymbol{i}}
\newcommand{\nuv}{\boldsymbol{\nu}}
\newcommand{\muv}{\boldsymbol{\mu}}
\newcommand{\lm}{\boldsymbol{\lambda}}
\newcommand{\Lm}{\boldsymbol{\Lambda}}
\newcommand{\opsi}{\overline{\psi}}
\renewcommand{\tan}{\textrm{tg}}
\renewcommand{\cot}{\textrm{ctg}}
\renewcommand{\sinh}{\textrm{sh}}
\renewcommand{\cosh}{\textrm{ch}}
\renewcommand{\tanh}{\textrm{th}}
\renewcommand{\coth}{\textrm{cth}}

\begin{abstract}
The tidal effects generated by a nonlinear gravitational wave are investigated in double-null v - u coordinates, as an exact solution of Einstein's field equations. The components $\xi^{v}$ and $\xi^{u}$ of the separation vector behave as in flat space but the transversal components $\xi^{x}$ and $\xi^{y}$ depend nonlinearly on $v$ through the Bessel and Neumann functions, far from the null surface $v = 0$. We show that the same results are obtained by means of the tetrad formalism.
  \end{abstract}

  \section{Introduction}
In General Relativity (GR), the equation of motion for a test particle is given by the geodesic equation which describe the worldline, i.e., the path through space and time of the particle. A geodesic is a curve of extremal length between any two points: its length remains unchanged to first order in small changes in the curve \cite{BS}. In flat space the freely falling particles maintain their separation but they do not in curved space. That leads to the geodesic deviation equation and show mathematically how the tidal forces of a gravitational field (which cause trajectories of neighboring particles to diverge) can be represented by curvature of spacetime. The affine parameter along a geodesic can be interpreted as proper time and is, thus, related to the reading of a clock that is transported with it \cite{DP}. 

To study the geodesic deviation phenomenon we have to find the variation of the separation vector between two neighboring geodesics, one of them being the reference-geodesic. The Fermi-normal coordinates are the best candidate for to investigate the second order effect of geodesic deviation. Manasse and Misner \cite{MM} computed the Schwarzschild (KS) metric in Fermi normal coordinates surrounding a radial geodesic, choosing a suitable orthonormal frame along the geodesic. They showed that the components of the Riemann tensor have the simplest form in those coordinates and all of them have the right dimension. Thus, normal coordinates provide a nice way by which a freely falling observer can report local experiments. 

Crispino et al. \cite{LC} analyzed the tidal forces produced in the Reissner-Nordstrom spacetime, finding some mismatches w.r.t. the KS geometry, depending of the mass-to-charge ratio. The radial components of the separation vector start decreasing inside the event horizon, unlike the KS case. Fleury \cite{PF} states that tidal forces experienced by ultrarelativistic particles in the direction of their motion are much smaller than those experienced orthogonally to their motion. Philipp and Puetzfeld \cite{PP} reconsidered the geodesic deviation in Newtonian gravitation and then determined relativistic effects within the theory of GR. Studying the first order deviation equation for orbits around the Earth, they uncovered artificial effects that are due to the linearized framework. The so-called Shirokov effect, rather than being a new feature of GR, they identified it as being the relict of the approximated description of a well-known perigee precession. Aldrovandi et al. \cite{APRV} assert that a physical gravitational wave cannot be represented by a solution to a linear wave equation. In their view, the effects produced by the 2nd order solution on free particles consist of nonlinear oscillations along the direction of propagation.

Our aim in this paper is to investigate the tidal effect produced on a test particle by a gravitational wave which is an exact solution of Einstein's field equations. We use the line-element from \cite{HC} in double-null coordinates, for to work with more simple form of the components of the Riemann tensor, compared to the Cartesian coordinates. Section 2 deals with the calculation of the components of the separation vector $\xi^{a}$ between two nearby geodesics , far away from the null surface $t - z = 0$, for a wave traveling along the z - axis. The only components with a nonlinear variation w.r.t. the affine parameter along the geodesic are the transversal components $\xi^{x}$ and $\xi^{y}$, expressed in terms of the Neumann and modified Bessel functions. A more convenient tetrad based formalism is applied in Section 3, for to obtain th deviation vector. Finally, in Section 4  we analize the implications of our results.

\section{Geodesic deviation in double-null coordinates}
Let us consider two freely falling particles which are very close to each other. Their worldlines are infinitesimally separated timelike geodesics \cite{PF}. We use the proper time $\tau$ to parameterize them and introduce the separation vector $\xi^{a}(\tau) = x^{a}(\tau) - y^{a}(\tau)$, which is orthogonal to the geodesics, i.e. $u_{a} \xi^{a} = 0$, where $u^{a}$ (with $u_{a} u^{a} = -1$) denotes the velocity 4-vector of one of the particles, tangent to the trajectory (here the index $a$ runs from $0$ to $3$). It is clear that $\xi^{a}$ represents their separation vector in the rest frame. The Jacobi equation
\begin{equation}
 \frac{D^{2}\xi^{a}}{d\tau^{2}}  +  R^{a}_{~bcd} u^{b}\xi^{c}u^{d} = 0
 \label{2.1}
 \end{equation}
gives the second covariant derivative of the deviation vector in terms of the Riemann curvature tensor, up to the linear order in the deviation and its first derivative (along the reference curve \cite{PP}). From (2.1) we infer that $\xi^{a}$ will have a nonlinear time dependence only if the geometry is curved, namely $R^{a}_{~bcd} \neq 0$. The capital $D$ in (2.1) signifies the covariant derivative and $R^{a}_{~bcd} = \partial_{c}\Gamma^{a}_{bd} - ...$.
We shall work in the local inertial frame at the point of the first geodesic where $\xi^{a}$ originates. In this coordinate system the coordinate distances are proper distances \cite{BS}. What is more, in that frame the covariant derivative acquires a simple form: in the local inertial frame the Christoffel symbols all vanish at that point, so the second derivative is just an ordinary one, w.r.t. the proper time. Therefore, Eq. (2.1) may now be written as \cite{BS}
 \begin{equation}
 \frac{d^{2}\xi^{a}}{d\tau^{2}}  +  R^{a}_{~bcd} u^{b}\xi^{c}u^{d} = 0,
 \label{2.2}
 \end{equation}
where $u^{a} = dx^{a}/d\tau$ is the velocity 4 - vector of the two particles.

We write down our proposed line-element corresponding to a nonlinear plane gravitational wave (GW) in double-null coordinates \cite{HC}, propagating along the z - coordinate
  \begin{equation}
  ds^{2} = - dv du + e^{-\frac{b}{\sqrt{v^{2} + b^{2}}}} dx^{2} + e^{\frac{b}{\sqrt{v^{2} + b^{2}}}} dy^{2} , 
 \label{2.3}
 \end{equation} 
 where $v = t - z$ is the retarded null coordinate, $u = t + z$ is the advanced null coordinate and $b$ is a constant length, taken here as the Planck length ($t, x, y, z$ denote the Cartesian coordinates). We take $v$ to play the role of time and the coordinates are $x^{0} = v,~ x^{1} = u,~ x^{2} = x,~x^{3} = y$. The geodesics in the spacetime (2.3) were investigated in \cite{HC}. We found there that $\dot{v} = dv/d\tau = 1$, so that one could take $v = \tau$ representing the proper time along the geodesics. We will specialize to the case $\dot{x} = u^{x} = 0,~\dot{y} = u^{y} = 0$, whence $u^{a} = (1, 1, 0, 0)$. \footnote{The fact that $\dot{u} = du/dv = 1$ comes from Eq. (4.6) from \cite{HC}, with $\dot{x} = \dot{y} = 0$.} Keeping in mind that  
 $u_{a} \xi^{a} = 0$, one finds that $\xi^{a} = (\xi^{v}, -\xi^{v}, \xi^{x}, \xi^{y}$), $\xi^{a}\xi_{a} = 1$.
 
 The nonzero components of the curvature tensor for (2.3) are given by \cite{HC}
  \begin{equation}
 \begin{split}
 R^{u}_{~xvx} = \frac{b}{2}\frac{bv^{2} - 2(2v^{2} - b^{2})\sqrt{v^{2} + b^{2}}}{(v^{2} + b^{2})^{3}} e^{-\frac{b}{\sqrt{v^{2} + b^{2}}}}\\
  R^{u}_{~yvy} = \frac{b}{2}\frac{bv^{2} + 2(2v^{2} - b^{2})\sqrt{v^{2} + b^{2}}}{(v^{2} + b^{2})^{3}} e^{\frac{b}{\sqrt{v^{2} + b^{2}}}}\\
   R^{x}_{~vvx} = \frac{b}{4}\frac{bv^{2} - 2(2v^{2} - b^{2})\sqrt{v^{2} + b^{2}}}{(v^{2} + b^{2})^{3}}\\
 R^{y}_{~vvy} = \frac{b}{4}\frac{bv^{2} + 2(2v^{2} - b^{2})\sqrt{v^{2} + b^{2}}}{(v^{2} + b^{2})^{3}},
 \label{2.4}
 \end{split}
 \end{equation}
 (the others are obtained by symmetry operations). By means of (2.4) it is an easy task to find that the components $\xi^{v}$ and  $\xi^{u}$ have a linear dependence on $v$. Hence
 \begin{equation}
 \frac{d^{2}\xi^{v}}{dv^{2}} = 0,~~~\frac{d^{2}\xi^{u}}{dv^{2}} = 0, 
 \label{2.5}
 \end{equation}
 As far as the other two components of $\xi^{a}$ are concerned, it is too complicate mathematically to work directly with the exact expression (2.4) of the curvature tensor, for to find the functional dependence of $\xi^{x}(v),~\xi^{y}(v)$. Therefore, we restrict ourselves to a more simple approach and consider that $v >> b$, i.e. the region far from the surface $v = b$, which is very close to the null surface $v = 0$ when $b$ is taken of the order of the Planck length. That restriction leads to the following equations for the other two components of  $\xi^{a}$
  \begin{equation}
 \frac{d^{2}\xi^{x}}{dv^{2}} + \frac{b}{v^{3}}\xi^{x} = 0,
 \label{2.6}
 \end{equation}
 and
   \begin{equation}
 \frac{d^{2}\xi^{y}}{dv^{2}} - \frac{b}{v^{3}}\xi^{y} = 0.
 \label{2.7}
 \end{equation}
 Let us deal firstly with Eq. (2.6). It is a linear, second order differential equation. Its solution may be expressed in terms of the Bessel functions \footnote{Thanks go to Viviana Ene and Luminita Cosma for helpful suggestions on the Bessel functions.}, one obtains, using \textit{Mathematica}
 \begin{equation}
 \xi^{x}(v) = C_{1} \sqrt{\frac{v}{b}} J_{1}\left(2\sqrt{\frac{b}{v}}\right) + C_{2} \sqrt{\frac{v}{b}} Y_{1}\left(2\sqrt{\frac{b}{v}}\right),
 \label{2.8}
 \end{equation}
 where $C_{1},~C_{2}$ are constants of integration, $J_{1}$ is the Bessel function of the first kind and index one and argument $2\sqrt{b/v}$ and $Y_{1}$ is the Neumann function of index one.  \cite{BK}. The first term gives us
  \begin{equation}
  \sqrt{\frac{v}{b}} J_{1}\left(2\sqrt{\frac{b}{v}}\right) = 1 - \frac{b}{2v} + \frac{b^{2}}{2v^{2}} - ...~~~,v >>b
 \label{2.9}
 \end{equation}
 But $Y_{1}$ is divergent when its argument tends to zero ($v >>b$) so that $\xi^{x}(v)$ becomes infinite when the retarded time $v \rightarrow \infty$. Now the two freely falling particles move away each other indefinitely.
 
 Let us look now for the solution of (2.7). That means to replace $b$ with $-b$ in (2.8). Therefore, we have
  \begin{equation}
 \xi^{y}(v) = C_{3} \sqrt{\frac{v}{b}} I_{1}\left(2\sqrt{\frac{b}{v}}\right) + C_{4} \sqrt{\frac{v}{b}} K_{1}\left(2\sqrt{\frac{b}{v}}\right),
 \label{2.10}
 \end{equation}
 where $C_{3},~C_{4}$ are constant of integration, $I_{1}$ is the modified Bessel function of index one and $K_{1}$ is the Macdonald function of index one. The first term of the r.h.s. of (2.10) can be written as a series expansion
    \begin{equation}
  \sqrt{\frac{v}{b}} I_{1}\left(2\sqrt{\frac{b}{v}}\right) = 1 + \frac{b}{2v} + \frac{b^{2}}{2v^{2}} - ...~~~,v >>b
 \label{2.11}
 \end{equation}
  But the second term depending on $K_{1}$ diverges when $b/v \rightarrow 0$. Hence, $|\xi^{y}(v)|$ increases indefinitely in this limit, having a similar behaviour with $\xi^{x}(v)$. That conclusion is also valid in flat space because, if we neglect the second term in (2.6) and (2.7), the ''no gravitation'' situation is recovered.
  
  We turn now our attention to the opposite case, namely $v << b$. Noting that in Eqs. (2.4) the above approximation leads to
   \begin{equation}
   R^{x}_{~vvx} = -R^{y}_{~vvy} = \frac{1}{2b^{2}},            
 \label{2.12}
 \end{equation}
 if we neglect $v^{2}/b^{2}$ w.r.t. unity. One obtains
    \begin{equation}
 \frac{d^{2}\xi^{x}}{dv^{2}} - \frac{1}{2b^{2}}\xi^{x} = 0,
 \label{2.13}
 \end{equation}
 and
   \begin{equation}
 \frac{d^{2}\xi^{y}}{dv^{2}} + \frac{1}{2b^{2}}\xi^{y} = 0.
 \label{2.14}
 \end{equation} 
  The solution for $\xi^{x}(v)$ is given by
  \begin{equation}
  \xi^{x}(v) =  C_{5} e^{-\frac{v}{\sqrt{2}b}} +  C_{6} e^{\frac{v}{\sqrt{2}b}},
 \label{2.15}
 \end{equation}
  with $C_{5}$ and $C_{6}$ - constants of integration. It is clear that, for $v << b$, the exponentials become unity and $\xi^{x}$ tends to a constant. In other words, there is no geodesic deviation along the x - direction. The solution for $\xi^{y}(v)$ appears as
   \begin{equation}
  \xi^{y}(v) =  C_{7} sin \frac{v}{\sqrt{2}b} +  C_{8} cos \frac{v}{\sqrt{2}b},
 \label{2.16}
 \end{equation} 
  with $C_{7}$ and $C_{8}$ - constants of integration. We have again $v << b$, whence $sin(v/b\sqrt{2}) \approx 0,~ cos(v/b\sqrt{2}) \approx 1$ and there is no geodesic deviation along the y - axis in this approximation.
  
  \section{Tetrad - based solution}
  Let us now compare the results obtained in Section 2 with those given by the tetrad formalism, applied for the metric (2.3). We use in this section the indices $a, b = 0, 1, 2, 3$ for the orthonormal basis and $\mu, \nu = 0, 1, 2, 3$ for the coordinate basis. We have 
    \begin{equation}
   \eta_{ab}e^{a}_{~\mu} e^{b}_{~\nu} = g_{\mu \nu},~~~ \eta^{ab}e_{a}^{~\mu} e_{b}^{~\nu} = g^{\mu \nu},
 \label{3.1}
 \end{equation}  
 where $\eta_{ab} = (-1, 1, 1, 1)$ is the Minkowski metric and $e^{a}_{~\mu}$ are the vierbeins. The only nonzero components of $g^{\mu \nu}$ are $g^{uv} = g^{vu} = -2,~g^{xx} = exp(b/\sqrt{v^{2} + b^{2}}),~g^{yy} = exp(-b/\sqrt{v^{2} + b^{2}})$. At each point we may define a local Lorentz frame using the orthonormal basis vectors $e_{a}$ which are not derived from any coordinate frame \cite{KL}, with the scalar product $e_{a}e_{b} = \eta_{ab}$ and $e_{a} = e_{a}^{\mu}e_{\mu}$ ( $e_{\mu}$ is the coordinate basis vector). 
 
 The local Lorentz frame at each point defines a family of ideal observers whose worldlines are the integral curves of the unit vector field $e_{0}$. The spacial unit vectors $e_{i} (i = 1, 2, 3)$ define the orthogonal space coordinates axes of a local laboratory frame, valid close to observer's trajectory \cite{KL}. In a freely falling frame with $u^{\mu} = (\dot{v}, \dot{u}, \dot{x}, \dot{y})$, we will choose $\dot{x} = \dot{y} = 0$, so that $\dot{u} = \dot{v}$ in this case (see \cite{HC}) and, therefore, $u^{\mu} = (\dot{v}, \dot{v}, 0, 0)$. For the metric (2.3), the tetrad may be written as \cite{BSV}
  \begin{equation}
  \begin{split}
  e^{0}_{~\mu} = \left(\frac{\dot{v}}{2}, \frac{1}{2\dot{v}}, 0, 0\right),~~~e^{1}_{~\mu} = \left(-\frac{\dot{v}}{2}, \frac{1}{2\dot{v}}, 0, 0\right),~~~\\
  e^{2}_{~\mu} = \left(0, 0, e^{-\frac{b}{2\sqrt{v^{2} + b^{2}}}}, 0\right)~~~e^{3}_{~\mu} = \left(0, 0, 0, e^{\frac{b}{2\sqrt{v^{2} + b^{2}}}}\right)
 \label{3.2}
 \end{split}
 \end{equation}  
 The contravariant components appear as
    \begin{equation}
  \begin{split}
  e_{0}^{~\mu} = \left(\frac{1}{\dot{v}}, \dot{v}, 0, 0\right),~~~,e_{1}^{~\mu} = \left(-\frac{1}{\dot{v}}, \dot{v}, 0, 0\right)~~~\\
  e_{2}^{~\mu} = \left(0, 0, e^{\frac{b}{2\sqrt{v^{2} + b^{2}}}}, 0\right)~~~e_{3}^{~\mu} = \left(0, 0, 0, e^{-\frac{b}{2\sqrt{v^{2} + b^{2}}}}\right)
 \label{3.3}
 \end{split}
 \end{equation}  
  But we have shown that, along any geodesic $\dot{v} = 1$ \cite{HC} and the expressions (3.2) and (3.3) simplify accordingly. Once we have the tetrads $e^{a}_{~\mu}$, the next step is to find the tidal force given by the geodesic deviation equation. For an observer in an orthonormal freely falling frame, the equation (2.2) should be written with ''caret''. However, we use only Latin indices for the tetrad indices and the Greek ones for the coordinate indices and so one will be no place for confusion. The separation vector $\xi^{a}$ joining two points on the neighbouring geodesics \cite{MS} is parameterized by some affine parameter which, as we have seen, may be chosen the retarded null coordinate $v$ in the local Lorentz frame. 
  
  The components of the Riemann tensor in the tetrad frame can be expressed as
   \begin{equation}
  R^{a}_{~bcd} = e^{a}_{~\mu}e^{~\nu}_{b}e^{~\sigma}_{c}e^{~\rho}_{d} R^{\mu}_{~\nu \sigma \rho}       
 \label{3.4}
 \end{equation} 
 in terms of its components in the coordinate basis. As in the previous Cartesian coordinates, we obtain from (2.2) and (3.4) that $\xi^{v}$ and $\xi^{u}$ depend linearly on $v$, 
  \begin{equation}
 \frac{d^{2}\xi^{v}}{dv^{2}} = 0,~~~\frac{d^{2}\xi^{u}}{dv^{2}} = 0, 
 \label{3.5}
 \end{equation} 
 as in flat space. The transversal components yield
   \begin{equation}
  \begin{split}
  \frac{d^{2}\xi^{x}}{dv^{2}} - \frac{b}{4}\frac{bv^{2} + 2(2v^{2} - b^{2})\sqrt{v^{2} + b^{2}}}{(v^{2} + b^{2})^{3}}\xi^{x} = 0,\\
  \frac{d^{2}\xi^{y}}{dv^{2}} - \frac{b}{4}\frac{bv^{2} - 2(2v^{2} - b^{2})\sqrt{v^{2} + b^{2}}}{(v^{2} + b^{2})^{3}} \xi^{y} = 0 . 
 \label{3.6}
 \end{split}
 \end{equation}  
  Let us noting that the above expressions are the same as that obtained from the last two equations (2.2), with reversed $\xi^{x}$ and $\xi^{y}$. That is the only difference w.r.t. the results obtained in double null coordinate basis, even though in the tetrad formalism there are more components of the curvature tensor that contribute to the Eqs. (3.6). Summarizing, the physical results are similar with those exhibited in Eqs. (2.8) and (2.10), using the same approximation.
  
  \section{Conclusions}
  In this paper we dealt with the geodesic deviation phenomenon applied to a curved geometry produced by a nonlinear GW. Based on the geodesic equations derived in a previous paper, we computed the separation vector between two nearby geodesics and found transversal effects only, in the region $v >> b$, namely far from the null surface $v = 0$. We studied also the tidal effects produced by the wave by means of an orthonormal basis and concluded that the separation vectors $\xi^{x}$ and $\xi^{y}$ have the same dependence on the retarded null coordinate $v$ as in the coordinate basis.

\end{document}